\documentclass[prd,twocolumn]{revtex4}
\usepackage{graphicx}
\usepackage{amssymb}
\usepackage{epstopdf}

\begin{document}

\title{Gravity with Perturbative Constraints: Dark 
Energy Without New Degrees of Freedom}
\date{\today}

\author{Alan Cooney}
\affiliation{Department of Physics, University of Arizona, 
Tucson, AZ 85721, USA}

\author{Simon DeDeo}
\affiliation{Kavli Institute for Cosmological Physics, 
University of Chicago, Chicago, IL 60637, USA \\
 \& \\ Institute for the Physics and Mathematics of the Universe, University of Tokyo, 
 Kashiwano-ha 5-1-5, Kashiwa-shi, Chiba 277-8582, Japan}

\author{Dimitrios Psaltis}
\affiliation{Departments of Astronomy and Physics, 
University of Arizona, Tucson, AZ 85721, USA}

\begin{abstract}
Major observational efforts in the coming decade are designed to probe
the equation of state of dark energy. Measuring a deviation of the
equation-of-state parameter $w$ from -1 would indicate a dark energy
that cannot be represented solely by a cosmological constant. While it
is commonly assumed that any implied modification to the $\Lambda$CDM
model amounts to the addition of new dynamical fields, we propose here a
framework for investigating whether or not such new fields are
required when cosmological observations are combined with a set of
minimal assumptions about the nature of gravitational physics. In our
approach, we treat the additional degrees of freedom as perturbatively
constrained and calculate a number of observable quantities, such as
the Hubble expansion rate and the cosmic acceleration, for a homogeneous
Universe. We show that current observations place our Universe within 
the perturbative validity of our framework and allow for the presence
of non-dynamical gravitational degrees of freedom at cosmological scales.
\end{abstract}

\maketitle 

\section{Introduction}
\label{intro}

Cosmological observations suggest the existence of dark matter. There
is, also, by now ample evidence that the Universe at late times is
accelerating.  Do either of these facts require the introduction of
new, dynamical degrees of freedom in the gravitational and matter sectors?

In the case of dark matter, the analysis of ``bullet cluster'' objects
such as 1E 0657-56~\cite{Clowe:2006p1018} and
MACSJ0025.4-1222~\cite{Bradac:2008p1023} comes as conclusive evidence
for the existence of new degrees of freedom independent of the visible
baryonic matter. While the exact nature of these new degrees of
freedom are still under debate, the simplest choice -- a single,
non-relativistic, dark matter fluid -- is the center of the
cosmological standard model.

Even if we choose to account for the galactic rotation curves with a
modification to gravity, it is difficult to do it without the addition
of extra degrees of freedom. A consequence of the Lovelock-Grigore
theorem~\cite{Lovelock:1970p948,Grigore:1992p984} is that most local,
classical, and covariant modifications to the gravitational action
lead to new fields~\cite{Skordis:2008p872}. Alternatives to dark
matter that have these properties -- such as
TeVeS~\cite{Bekenstein:2004p1081} and STVG~\cite{Moffat:2006p1065} --
themselves have new fields as a consequence of the Lovelock-Grigore
theorem.

Today, cold dark matter with a cosmological constant ($\Lambda$CDM)
satisfies all current observational
constraints~\cite{2008arXiv0803.0547K}. However, a number of
theoretical issues~\cite{Weinberg:1989p1168} have prompted
observational campaigns to search for deviations from the predicted
$\Lambda$CDM behavior~\cite{Albrecht:2006p1174}. These observations
will measure the equation-of-state parameter $w$ of the dark energy
and search for deviations from the value $w=-1$ that corresponds to a
cosmological constant. Should their results force us to abandon
$\Lambda$CDM, we will face a problem for dark energy similar to that
we faced for dark matter: do such observations require new dynamical
degrees of freedom?  So far, the two major approaches to dark energy
modeling that predict values of $w\ne -1$,
quintessence~\cite{ArmendarizPicon:2000p1215} and $f(R)$
gravity~\cite{Carroll:2004p1253}, incorporate new dynamical degrees of
freedom, which can be cast, in both cases, in the form of a single
scalar field. 

A number of attempts have been made for modifying General Relativity
in a way that generates cosmologies with $w\ne -1$ but without
introducing new dynamical fields.  Modified source
gravity~\cite{Carroll:2006p871}, based on insights from the Palatini
formulation of the Einstein equations~\cite{Flanagan:2004p1310} forces
the (Einstein-frame) scalar field to be non-dynamical by erasing the
kinetic term. Cuscuton
cosmology~\cite{Afshordi:2007p59,Afshordi:2007p54} also modifies the
kinetic term; it appears to have new dynamical degrees of freedom but
these are frozen out at the perturbative level. Non-local~\cite{Barvinsky:2003p1420,Soussa:2003p1374}
 or holographic~\cite{Larsen:2002p4709} theories may
also avoid the introduction of new degrees of
freedom.  Albeit
ambitious, these attempts can face serious
problems~\cite{Iglesias:2007p869}.

In the opposite extreme, because of the Lovelock-Grigore theorem, the
theoretical pressures on purely phenomenological models that deviate
from $\Lambda$CDM but that do not introduce new dynamical degrees of
freedom are strong. The fact that observations favoring dark energy
are made on relativistic, Hubble scales makes it even harder to avoid
the consequence of the Lovelock-Grigore theorem, while remaining
internally consistent. Requiring general covariance from the start,
for example, prevents one from even phrasing many minimal theories the
way MOND did on galactic scales~\cite{Milgrom:1983p1141}. In general,
while many theories similar or equivalent to Brans-Dicke~\cite{BD61} may,
in the limit of small post-Newtonian corrections, appear only to modify the General Relativistic
equations of motion, the new scalars such theories introduce appear as truly independent degrees
of freedom in the relativistic regime -- for example, on scales approaching the horizon size.

Approaches, however, that tie together two particular sets of observations
-- homogeneous expansion and linear structure formation -- in a
phenomenological but self-consistent fashion have also been recently
explored. For example, the parametrized post-Friedmann
framework~\cite{Hu:2007p1209} makes minimal assumptions about the
underlying physics, requiring only causality, metric structure, and
energy-momentum conservation. The Cardassian
model~\cite{2002PhLB..540....1F} introduces a power-law term in the
Friedmann equation, which leads to accelerated expansion even in a
matter dominated universe. The approach of
Ref.~\cite{Skordis:2008p872} is similar in spirit, and explicitly
shows how a subset of modifications will introduce new degrees of
freedom. Although these approaches are general and simple to use in
connection to observational data, they can not, by their very nature,
generate predictions for other astrophysical settings -- the solar
system, for example, or nearby (non-cosmological) event horizons.
As a result, they can not guarantee that dynamical degrees of freedom are
not required for consistency in these other situations.

We follow here a different approach intermediate to the two outlined
above, which we call Gravity with Perturbative Constraints (see also
Ref.~\cite{2008PhRvD..78f4013D} for an initial discussion of cosmology
with perturbative constraints).  In contrast to the fundamental
approach, we do not propose a basic theory but rather remain largely
neutral on the details of how a particular set of observations are
generated. In contrast to the phenomenological approach, our models
have a wide range of applicability and one can, for example, ask
whether the cosmological predictions that do not add extra degrees of
freedom are consistent with observations in the solar system or of
compact objects.

Our aim is to cover as large a class as possible of modifications to
gravity that do not require new degrees of freedom. Within our
framework, it is possible in principle to incorporate all such
modifications that are covariant, metric (in the sense of obeying the
Einstein equivalence principle), and conserve energy-momentum.  The
price we will pay for such generality is that we must remain
perturbatively close to General Relativity.  While we discuss in
detail the exact conditions for perturbative validity below, we
caution here that this valid- ity does {\em not} require weak fields
(as defined with reference to either the Minkowski or homogenous FRW
metrics as the background).

Our method of generating such models is known in the literature as
that of perturbative constraints~\cite{Simon:1990p1512}; it is an
alternative way of associating, in a physically unambiguous fashion, a
general action of the gravitational field with a field equation that is
of second order. Developed to deal with approximate local actions derived from
fundamentally non-local theories, it has a wide range of applicability
and does not rely on the existence of an underlying non-locality.

The method of perturbative constraints we use here allows us to
retain many of the important features of the action of the
gravitational field. In particular, it allows us to retain the
translational symmetries that, under Noether's principle, lead to
energy-momentum conservation, as well as enforcing diffeomorphism
invariance -- the invariance of physical law under changes of
coordinates. It does so without introducing new degrees of freedom,
while at the same time producing behavior that differs from the
General Relativistic case, and allowing for a consistent set of
predictions across a wide variety of spacetimes.

\section{Gravity with Perturbative Constraints}

In this paper, we specialize to the case of a flat, homogenous FRW metric,
\begin{equation}
\small{ds^2 = -dt^2 + a^2(t)\left[dr^2+ 
r^2\left( d\theta^2 + \sin^2\theta d\phi^2\right)\right]} \;.
\end{equation}
It is important to emphasize here, however, that because we will be
modifying gravity in ways that affect cosmological expansion, standard
relationships between, e.g., the energy density of various components
and the rate of expansion that hold in a flat Universe within General
Relativity will not necessarily hold here.

We start with an action for the gravitational field that is a general
function $f(R)$ of the Ricci scalar curvature.  The field equation for
this general theory is \begin{equation} f' R_{\mu \nu}
-\frac{1}{2}g_{\mu \nu}f + g_{\mu \nu} \square f' - \nabla_{\mu}
\nabla_{\nu} f' = 8\pi G T_{\mu \nu}\;,
 \end{equation} 
where $f'=\frac{\partial f}{\partial R}$. For a flat FRW spacetime, the
Ricci scalar curvature is
\begin{equation}
R= 6\left[ \frac{\ddot{a}}{a}+ \left(\frac{\dot{a}}{a}\right)^2 \right]\;.
\end{equation}
The above field equation obeys the Bianchi identity and hence
$\nabla^{\mu}T_{\mu \nu}=0$, which results in the usual conservation
law
\begin{equation}
\dot{\rho}_i + 3\frac{\dot{a}}{a}\left( \rho_i + P_i\right) = 0 \;.
\label{eq:cons}
\end{equation}
Here $\rho_i$ and $P_i$ are the energy density and pressure,
respectively of the $i^{\mbox{\scriptsize{th}}}$ constituent of the
Universe. In this paper, we will consider Lagrangian densities of the
form
\begin{equation}
 f(R)=R-2\Lambda +\frac{\mu^{2(n+1)}}{R^n} \; ,
\end{equation}
where n and $\mu$ are free parameters and the last term in the action
is considered only at the perturbative level. The case $\Lambda = 0$,
$n=1$ was first discussed in Ref.~\cite{2008PhRvD..78f4013D}.

Henceforth we shall adopt geometerized units where $8\pi G = 1$.
Following standard procedure, we evaluate the $(\mu, \nu) = (0,0)$
component of the field equation as well as its trace and obtain
\begin{equation}
\label{00}
 -3f'\frac{\ddot{a}}{a}+\frac{1}{2}f+3f''\dot{R}\frac{\dot{a}}{a}= \rho 
 \end{equation}
and
\begin{equation}
\label{tr}
\small{Rf' -2f-3\left[ f''\ddot{R} + \dot{R}^2 f''' +3\frac{\dot{a}}{a}\dot{R}f'' \right]=\left( -\rho +3P\right)} \;.
\end{equation}
In this last equation, $\rho$ and $P$ denote the total density and
pressure of all constituents of the Universe.

When the function $f(R)$ is non linear in the Ricci scalar curvature,
Eq.~\ref{00} is not a first-order ordinary differential
equation in time, as is the standard Friedmann equation.  Indeed, as
discussed earlier, a non-linear Lagrangian introduces additional
degrees of freedom to the gravitational field. The key feature of
gravity with perturbative constraints is a consistent way of treating these additional
degrees of freedom as fixed and not allowing them to evolve
independently~\cite{EW89,Simon:1990p1512,2008PhRvD..78f4013D}.

In practice, we combine Eq.~\ref{00} with the conservation
law of Eq.~\ref{eq:cons}, and obtain a single differential equation for the
scale factor that we formally denote by
\begin{equation}
{\cal G}(\dot{a},a)+\mu^4{\cal
F}(\dddot{a},\ddot{a},\dot{a},a)=0\;.
\label{eq:all_deg_freed}
\end{equation}
Solving this equation with $\mu^4=0$, i.e.,
\begin{equation}
{\cal G}(\dot{a}^{(0)},a^{(0)})=0\;.
\end{equation}
and requiring the scale factor to be unity at the present epoch, we
obtain the General Relativistic solution, which we denote by
$a^{(0)}(t)$.  We then use this solution to evaluate the derivatives
that correspond to all the new degrees of freedom in
Eq.~(\ref{eq:all_deg_freed}), i.e.,
\begin{equation}
{\cal G}(\dot{a},a)+\mu^4{\cal F}(\dddot{a}^{(0)},\ddot{a}^{(0)},
\dot{a}^{(0)},a^{(0)})=0\;.
\end{equation}
This last equation is again an ordinary differential equation that is
of first order in time and we solve it requiring the scale factor to be
unity in the present epoch.

Inspection of the field equations or of the Lagrangian action shows
that we can expand the final solution for the scale factor in orders
of the parameter $\mu^{2(n+1)}/R_0^{n+1}$, where
$R_0=\rho_{m,0}+4\Lambda$ is the unperturbed Ricci scalar curvature of the
Universe in the present epoch. For the case with $n=1$, we can write
the solution for the scale factor in the form
\begin{eqnarray}
a(t) &=& a^{(0)}(t)\left[1+
\frac{\mu^4}{(\rho_{m,0}+4\Lambda)^2} a^{(4)}(t) \right.\nonumber\\
&&\qquad\qquad+\left.
\frac{\mu^8}{(\rho_{m,0}+4\Lambda)^4} a^{(8)}(t) + \ldots  \right]\;.
\label{eq:orders}
\end{eqnarray}
Note that, although the modified Friedmann equation includes only
terms of order $\mu^4$, its solution has higher-order terms because of
the non-linearity of the differential equation. Consistency requires
that terms of order $\mu^8$ and higher are negligible and can,
therefore, be dropped.

When discussing our results, we will occasionally express them in
terms of the contributions of matter and of the cosmological constant
to the energy density of the Universe, which are given by
\begin{equation}
\Omega_{\rm m}\equiv \frac{\rho_m}{3H^2}
\label{eq:Omegam}
\end{equation}
and
\begin{equation}
\Omega_\Lambda\equiv\frac{\Lambda}{3H^2}\;,
\label{eq:Omegal}
\end{equation}
respectively, where $H\equiv (\dot{a}/{a})$.  Finally, in order to
compare our results to the usual parametrization of the dark energy
equation of state, we will also write the Friedmann equation in the
XCDM form~\cite{2003RvMP...75..559P} as
\begin{equation}
3\left(\frac{\dot{a}}{a}\right)^2 = \rho_{\rm m} + \rho_{\rm X}\;,
\label{eq:XCDM}
\end{equation}
and the equation for acceleration as
\begin{equation}
6\frac{\ddot{a}}{a} = -\rho_m-(1+3w_{\rm X})\rho_{\rm X}\;,
\end{equation}
where $\rho_X$ is the energy density of an equivalent ``dark energy''
component with an equation of state $P_{\rm X}=w_X\rho_{\rm X}$. Note
that this term includes contributions from both the cosmological
constant and by the perturbative term that modifies the Einstein
equations.

\subsection{Case $1$: $\Lambda=0$, $n=1$}

This situation corresponds to a Universe with zero cosmological
constant and $f(R) = R+\mu^{4}/R$. Since we are interested in
deviations in the matter dominated regime, we will only consider
matter with $P=0$. In this case, the $(\mu, \nu) = (0,0)$ component of
the field equation and its trace become
\begin{equation}
3 \left(\frac{\dot{a}}{a}\right)^2+\frac{3\mu^4}{R^2}
\left[ \frac{R}{6}+\frac{\ddot{a}}{a}+2\left(\frac{\dot{a}}{a}\right)
\frac{\dot{R}}{R}\right] = \rho_{m}
\label{eq:Hubb}
 \end{equation}
and
\begin{equation}
R +\frac{3\mu^4}{R^2}\left[ R+ 2\frac{\ddot{R}}{R}-
6\left( \frac{\dot{R}}{R}\right)^2 +6\left(\frac{\dot{a}}{a}\right)
\frac{\dot{R}}{R}\right] = \rho_{m} \;,
\end{equation}
respectively.

When $\mu=0$, the equation for the scale factor describes the familiar
matter-dominated Einstein cosmology, for which
\begin{equation}
\label{matdom0}
a^{(0)}(t)= \left(\frac{\sqrt{3\rho_{m,0}}t}{2}\right)^{\frac{2}{3}} \;.
\end{equation}
As a result 
\begin{equation}
\left(H^{(0)}\right)^2\equiv \left(\frac{\dot{a}^{(0)}}{a^{(0)}}\right)^2=\frac{ \rho_{m,0}}{3\left(a^{(0)}\right)^3}
\label{H0zero}
\end{equation}
and 
\begin{equation}
R^{(0)} \equiv 6\left[ \frac{\ddot{a}^{(0)}}{a^{(0)}}+ \left(H^{(0)}\right)^2\right]= \frac{ \rho_{m,0}}{\left(a^{(0)}\right)^3}\;.
\label{Rzero}
\end{equation}
Now to order $\mu^4$ Eq.~(\ref{eq:Hubb}) gives
\begin{eqnarray}
2H^{(0)}\dot{a}^{(4)} + &&\frac{\mu^4}{(R^{(0)})^{2}}\left( \frac{R^{(0)}}{6} + \frac{\ddot{a}^{(0)}}{a^{(0)}}+2\frac{\dot{R}^{(0)}}{R^{(0)}} H^{(0)}\right) \nonumber \\
&&= -a^{(4)}\frac{\rho_{m,0}}{\left(a^{(0)}\right)^3}\;.
\end{eqnarray}
Solving for $a^{(4)}(t)$ using Eq.~(\ref{H0zero}) and~(\ref{Rzero}) the contribution to order $\mu^4$ is
\begin{equation}
a^{(4)}(t) =\frac{9\rho_{m,0}^2t^4}{40}\;.
\end{equation}
We then perform the same expansion up to order $\mu^8$. The contribution is
\begin{equation}
 a^{(8)}(t) = -\frac{18819\rho_{m,0}^4t^8}{1600}\;,
\end{equation}
As a boundary condition, we require that at the present
epoch, which we denote by $t_0$, the scale factor is unity. Thus, as we
evaluate $a(t)$ to higher order in $\mu$, we get corrections to the
value of $t_0$ from General Relativity, i.e.,
\begin{equation}
t_0 = \frac{2}{\sqrt{3\rho_{m,0}}}
\left(1 -\frac{3}{5}\frac{\mu^4}{\rho_{m,0}^2}+ 
\frac{115}{2}\frac{\mu^8}{\rho_{m,0}^4}+ \ldots \right)
\end{equation}

As discussed earlier, it is practically impossible to check the
convergence of our method without knowing the form of the Lagrangian
action to higher order in the parameter $\mu$. We can provide,
however, a necessary condition for convergence in calculating an
observable quantity, by requiring that the correction to the value of
this observable calculated to order $\mu^8$ is not larger than the
correction calculated to order $\mu^4$.

As an example, we calculate the value of the Hubble constant at the
present epoch, $H_0\equiv H(t=t_0)$, as
\begin{equation}
H_0=\sqrt{\frac{\rho_{m,0}}{3}}\left( 1+3\frac{\mu^4}{\rho_{m,0}^2}-\frac{1017}{2}\frac{\mu^8}{\rho_{m,0}^4}+ \ldots \right)\;.
\end{equation}
Similarly we evaluate the deceleration parameter $q_0\equiv
-\ddot{a}a/\dot{a}^2$ at the present epoch as
\begin{equation}
q_0=\frac{1}{2}\left(1-36\frac{\mu^4}{\rho_{m,0}^2} +
12312\frac{\mu^8}{\rho_{m,0}^4}+ \ldots \right)\;.
\end{equation}
Requiring the correction to order $\mu^8$ in the expression for the
Hubble constant to be smaller than the correction to order $\mu^4$
results in the bound $\mu^4/\rho_{\rm m,0}^2 < 6/1017$, whereas the
same requirement for the deceleration parameter leads to
$\mu^4/\rho_{\rm m,0}^2 < 36/12312$.  Both of these constraints at the
present epoch are very similar and require $\mu^4$ to have very small
values, which cannot lead to an accelerated expansion within the
limits of perturbative validity.

An additional problem of using a cosmology with $n=1$ and $\Lambda=0$
to account for the observations can be seen by writing the Friedmann
equation to order $\mu^4$ (as in~\cite{2008PhRvD..78f4013D}), i.e.,
\begin{equation}
3\left(\frac{\dot{a}}{a}\right)^2 
= \rho_{\rm m}+ \frac{6 \mu^4}{\rho_{\rm m}}\;.
\end{equation}
Transforming this equation in the XCDM form
requires that
\begin{equation}
\rho_X = \frac{6\mu^4}{\rho_{\rm m}}\;.
\end{equation}
Clearly the matter-dominated perturbative gravity theory to order
$\mu^4$ behaves as a dark energy fluid with $w_{\rm X}=-2$ which is 
inconsistent with the combined WMAP and BAO data~\cite{2008arXiv0803.0547K}.


\subsection{Case $2$: $\Lambda \ne0$, $n=1$}

We have found that, while the case with $n=1$ and $\Lambda=0$ may
generate accelerating solutions, these solutions lie outside the range
where the perturbative expansion to the action is valid. Intuitively,
this comes from the fact that by the time acceleration has begun, the
matter density is low enough that the zeroth-order relation,
$R_0=\rho_{\rm m,0}$, implies that the contribution of the additional
term, $\mu^4/R$, is too large to be treated as a small perturbation to
the standard gravitational action. This is not the case, however, in a
Universe with a non-zero cosmological constant, as we will discuss
below.

In the case of a Universe with a non-zero cosmological constant, the
$(\mu ,\nu)=(0,0)$ component of the field equation and the trace are
\begin{equation}
\label{lamcase}
3\left(\frac{\dot{a}}{a}\right)^2+\frac{3\mu^4}{R^2}\left[
\frac{R}{6}+\frac{\ddot{a}}{a}+2\left(\frac{\dot{a}}{a}\right)
\frac{\dot{R}}{R}\right]
= \rho_{\rm m} +\Lambda\;,
\end{equation} 
and
\begin{equation}
R +\frac{3\mu^4}{R^2}\left[ R+ 2\frac{\ddot{R}}{R}
-6\left( \frac{\dot{R}}{R}\right)^2 
+6\left(\frac{\dot{a}}{a}\right)\frac{\dot{R}}{R}\right] 
= \rho_{\rm m} + 4 \Lambda\;,
\end{equation}
respectively.

We will again seek a perturbative solution to this equation of the
form~(\ref{eq:orders}). When $\mu=0$, the Friedmann equation is the
same as for $\Lambda$CDM, i.e.,
\begin{equation}
a^{(0)}(t)= \left[ \sqrt{\frac{\rho_{m,0}}{\Lambda}} 
\sinh \left(\frac{\sqrt{3\Lambda}}{2}t\right) \right]^\frac{2}{3}\;.
\end{equation}
The time at the present epoch, i.e., the one for which the
$\Lambda$CDM solution gives a scale factor of unity, is
\begin{equation}
t_0 = \frac{2 }{\sqrt{3\Lambda}}\sinh^{-1}
\left(\sqrt{\frac{\Lambda}{\rho_{m,0}}}\right)\;.
\end{equation}

\begin{figure}[t]
\includegraphics[width= 85mm]{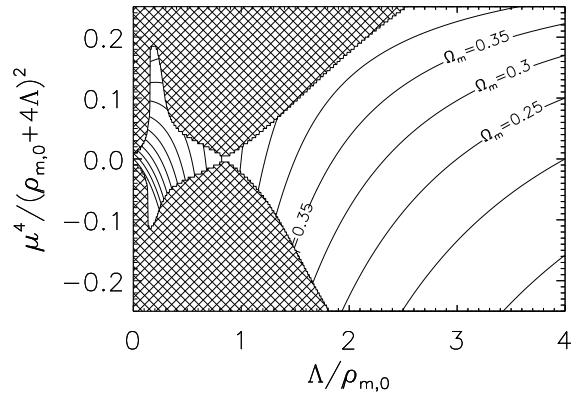}
\caption{Contours of constant values of the parameter $\Omega_{\rm m}$ at
the present epoch, for different values of the cosmological constant
$\Lambda$ and the parameter $\mu^4$ of the action of the gravitational
field. The normalization constant $\rho_{\rm m,0}$ is the density of
matter at the present epoch.  The hatched area shows the region of the
parameter space where the calculation of $\Omega_{\rm m}$ is not
perturbatively valid.
\label{fig:Omegam}}
\end{figure}

\begin{figure}[t]
\includegraphics[width= 85mm]{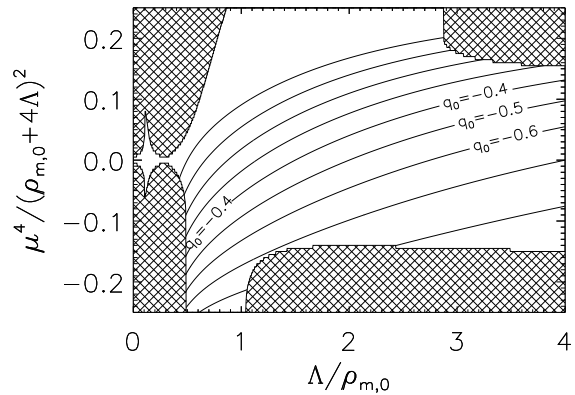}
\caption{Same as Figure~\ref{fig:Omegam} for the deceleration parameter
$q_0$ at the present epoch. Only negative values of $q_0$ are shown,
which correspond to Universes that are currently accelerating.
\label{fig:q0}}
\end{figure}

At orders $\mu^4$ and $\mu^8$, the solutions of the equation for the
scale factor, $a^{(4)}(t)$ and $a^{(8)}(t)$, respectively, as well as
the time $t_0$ that corresponds to the present epoch can be found in
closed form but are too long to be displayed here (we present the
solution to order $\mu^4$ in Appendix~A). Instead, we write the
Friedmann equation to order $\mu^4$ 
\begin{eqnarray}
\label{hubl}
3\left(\frac{\dot{a}}{a}\right)^2 &=& \rho_m+ \Lambda-
\frac{\mu^4}{(\rho_m+4\Lambda)^2}\nonumber\\
&&\qquad
\left[3 \Lambda - 6 (\rho_m+\Lambda)
\left(\frac{\rho_m}{\rho_m+4\Lambda} \right) \right] \;.
\end{eqnarray}
and the acceleration equation to the same order 
\begin{eqnarray}
\label{accl}
6\frac{\ddot{a}}{a} &=& -\rho_m+2\Lambda -\frac{\mu^4}{(\rho_m+4\Lambda)^2}
\nonumber\\
&&\qquad
\left[ 6\Lambda +24(\rho_m+\Lambda)
\left( \frac{\rho_m}{\rho_m + 4\Lambda}\right)\right.\nonumber\\
&&\qquad\left.
 -54(\rho_m+\Lambda)\left( \frac{\rho_m}{\rho_m + 4\Lambda}\right)^2\right]\;.
\end{eqnarray}
These two equations allow us to calculate to order $\mu^4$ two
observable quantities at the present epoch, as a function of the
parameters $\mu^4$ and $\Lambda$: the Hubble constant, or equivalently
the parameter $\Omega_{\rm m,0}$, and the deceleration parameter
$q_0$. We show the result of this calculation in Fig.~\ref{fig:Omegam}
and \ref{fig:q0} respectively, together with the regions where
the calculation of each observable is not perturbatively valid.

In both figures, the horizontal axis, i.e., $\mu^4/(\rho_{\rm
m,0}+4\Lambda^2)=0$, corresponds to a General Relativistic Universe
with a cosmological constant. The WMAP results, interpreted within
General Relativity, correspond to a Universe with $\Lambda/\rho_{\rm
m,0}\simeq 2.65$~\cite{2008arXiv0803.0547K}. Keeping this value
constant and moving towards positive values of $\mu^4/(\rho_{\rm
m,0}+4\Lambda^2)$, the predicted value of $\Omega_{\rm m,0}$
increases, because the perturbative term in the modified Friedman
equation contributes negatively to the rate of expansion of the
universe, and the deceleration parameter decreases for the same
reason. The opposite is true when the parameter $\mu^4/(\rho_{\rm
m,0}+4\Lambda^2)$ becomes increasingly negative.

The hatched areas in both figures show the regions of the parameter
space where the corrections to each observable calculated to order
$\mu^8$ are not negligible and hence the solution is no longer of the
same order as the field equation. This is the criterion we have
discussed earlier in determining the perturbative validity of our
calculations. The particular shapes of the hatched regions are
determined by the dependence of the perturbative terms in
Eqs.~\ref{hubl} and \ref{accl}, both of which can be negative,
positive, or even zero for different values of the ratio
$\Lambda/\rho_{\rm m,0}$. In both figures, however, the excluded
regions lie far from the parameters of the General Relativistic
Universe that are consistent with the WMAP data, suggesting that small
potential deviations from the General Relativistic predictions can be
modeled successfuly within our framework.

\begin{figure}[t]
\includegraphics[width= 85mm]{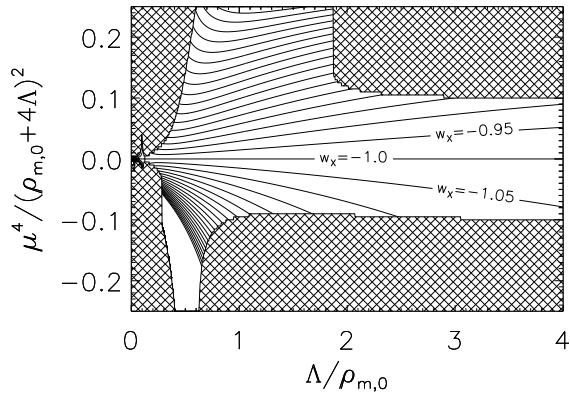}
\caption{Same as Figure~\ref{fig:Omegam} for the equivalent 
equation-of-state parameter $w_{\rm X}$ of a XCDM model with the same
cosmological expansion at the present epoch.
\label{fig:w0}}
\end{figure}

Finally, we investigate the parameter $w_{\rm X}$ at the present epoch
for an equivalent XCDM cosmology that would result in the same
cosmological expansion rate and acceleration as that predicted by the
last two equations. We calculate the equivalent equation-of-state
parameter by comparing the last two equations to those of the XCDM
framework. The result is, to order $\mu^4$,
\begin{eqnarray}
\label{wnumbs}
w_{\rm X} &\sim& -1+\frac{\mu^4}{(\rho_m+4\Lambda)^2} 
\left(\frac{\rho_m + \Lambda}{\Lambda}\right) 
\nonumber\\
&& \left[ 12\left( \frac{\rho_m}{\rho_m + 4\Lambda}\right)  
-18\left( \frac{\rho_m}{\rho_m + 4\Lambda}\right) ^2\right]\;.
\end{eqnarray}
In Fig.~\ref{fig:w0}, we show contours of constant values of the
parameter $w_{\rm X}$. The current observed limits $-0.14<w_{\rm
X}<0.12$ on the equation-of-state parameter~\cite{2008arXiv0803.0547K}
suggest that deviations from the $\Lambda$CDM model that do not
introduce new degrees of freedom fall within the perturbative limits
of our approach and cannot be ruled out {\em a priori\/}.

\subsection{Case $3$: $\Lambda=0$, $n \ne1$}
\label{nne1}

Finally, we examine a more general class of theories with $n \ne 1$
that are nevertheless matter dominated. We focus here on the case
$\Lambda=0$ in order to isolate the expected dependence of our previous
results on the value of the parameter $n$. 

In this case, the $(\mu ,\nu)=(0,0)$ field equation becomes
\begin{equation}
 3\left(\frac{\dot{a}}{a}\right)^2+\frac{3\mu^{2(n+1)}}{R^{(n+1)}}
\left[ \frac{R}{6}+n\frac{\ddot{a}}{a}+n(n+1)\left(\frac{\dot{a}}{a}\right)
\frac{\dot{R}}{R}\right] = \rho_{\rm m}
\end{equation}
and its solution can be expanded again in terms of the perturbative
parameter $\mu^2/R_0$. The constrain on this parameter for
perturbative validity becomes increasingly less restrictive as the
exponent $n$ decreases. 

As with Case 1, we can write the Friedmann equation to order
$\mu^{2(n+1)}$ and obtain
\begin{equation}
3\left(\frac{\dot{a}}{a}\right)^2
 = \rho_{\rm m}+ \frac{\mu^{2(n+1)}}{2\rho_{\rm m}^n}
\left[6n(n+1)+n-1\right]\;.
\end{equation}
At the same order the acceleration equation is given by
\begin{equation}
6\frac{\ddot{a}}{a} = -\rho_{\rm m} + \frac{(2+3n)\mu^{2(n+1)}}
{2\rho_{\rm m}^n}\left[6n(n+1)+n-1\right]\;.
\end{equation}

If we write the Friedmann equations in the flat XCDM form,
Eq.~(\ref{eq:XCDM}), in the present epoch, we obtain for the equivalent
equation-of-state parameter,
\begin{equation}
w_{\rm X} = -(n+1)\;.
\end{equation}
As a result, increasing the value of the exponent $n$ makes the
equation-of-state parameter $w_{\rm X}$ more negative.

\section{Conclusions}

We have presented a formalism -- Gravity with Perturbative Constraints
-- that describes deviations from General Relativistic predictions at
cosmological scales without introducing new degrees of freedom. Our
formalism preserves diffeomorphism invariance, the Einstein equivalence
principle, and energy-momentum conservation. Because it is based on
modifications of the Lagrangian of the gravitational field, it allows
us to link a wide variety of observations -- from compact objects and
solar system phenomena, to non-linear clustering and linear structure
formation -- in order to test for the consistency of assuming no new
gravitational degrees of freedom.

Our method is quite different from the Palatini formulation of 1/R gravity discussed by, for example, Ref.~
\cite{v03}, and expanded on as ``modified source gravity'' in Ref.~\cite{Carroll:2006p871}. Since both our methods and those related to Palatini formulations, however, eliminate the extra degree of freedom associated with modifying General Relativity it is worth comparing some of their features. In particular, because we are able to treat our system perturbatively -- and, explicitly, because our predictions for densities $\rho \gtrsim\rho_{\mathrm{crit}}$ do not demand a particular behavior when $\rho\ll\rho_{\mathrm{crit}}$ -- we do not run into the issues found by Ref.~\cite{Flanagan:2004p1310}, where the microscopic discreteness expected for ordinary baryons and dark matter particles no longers ``averages out'' correctly to return the homogenous Friedmann equations.

Put explicitly, Ref.~\cite{Flanagan:2004p1310} noted first that the microscopic discreteness of matter did not strongly affect the averaged metric in General Relativity: \emph{i.e.}, when taking the matter field to be composed of discrete particles (small but still larger than their Schwartzschild radius) the gravitational fields relative to the background are still small and $g_{\mu\nu}$ can be consistently be replaced by $\langle g_{\mu\nu} \rangle$ plus a small, linear perturbation. This behavior was then contrasted with the Palatini $1/R$ formulation, where the $\Phi$ field, a function of the matter density that appears as a modification to the Einstein equation, has strong variations from point to point since the behavior of $\Phi$ is strongly non-linear in the range $\rho\in[0,\rho_{\mathrm{crit}}]$ regardless of the length scale of the perturbation.

In the case of gravity with perturbative constraints, however, describing the behavior of the equations of motion when $\rho\ll\rho_{\mathrm{crit}}$ may involve increasingly higher order terms in the approximate equations of motion, and -- likely for many, though perhaps not all, models where the first terms in the equations of motion are generated by $1/R^n$ cases discussed in this paper -- eventually goes outside the radius of convergence of the series. That our theory can not in all cases make predictions for atomic-scale physics is a feature of the perturbative approach; in contrast to Palatini $1/R$, the behavior at $\rho \gtrsim\rho_{\mathrm{crit}}$ decouples from these more difficult questions and no particular atomic-scale behavior is required by us for consistency with larger scales.

In this first approach to the problem, we chose to look for
modifications of the Einstein-Hilbert action that involve the addition
of terms proportional to an inverse power of the Ricci scalar
curvature. We were lead to this choice by our goal to describe
potential deviations at cosmological scales, without affecting
significantly the behavior of gravity in the solar system.  One might
object that our particular choice of parametrization -- a two
dimensional space, ($\Lambda$, $n$) -- too-drastically narrows down
the possible models considered. In a sense, this is a problem with any
attempt to cover a space of functions with a finite number of
parameters.

Our approach, however, follows the general direction of testing for
modifications of gravity in the solar system and in other
astrophysical systems by using the Parametrized Post-Newtonian (PPN)
framework \cite{10887}.  In a similar fashion to our work, the PPN
framework provided a way to link together different observations in
various astrophysical settings in a consistent and physically rigorous
way.  While the PPN model itself required a certain degree of
arbitrariness -- the functional forms of the potentials -- it turned
out that the original choices of these functional
forms~\cite{Will:1972p1578} were sufficient to cover nearly every
theory proposed in the following thirty-seven years. Only in a few cases
have failures of PPN been due to an overly restrictive choice of
parameters (see, \emph{e.g.}~\cite{Alexander:2007p1595}). The usefulness of the framework
we propose here will depend, of course, on its ability to capture the
behavior of plausible gravitational theories that go beyond General
Relativity.

The particular set of terms that we included in the Lagrangian of the
gravitational field allow us also to study in a more systematic way
other phenomenological cosmologies that produce accelerating
expansion. For example, the set of parameters discussed in
Sec.~\ref{nne1}, i.e., $\Lambda=0$ and $n\ne 1$, provide a homogenous cosmology
identical to the Cardassian model of
Ref.~\cite{2002PhLB..540....1F}. Within our formalism, however, we can
also make predictions for the formation of structure in this
cosmology, by taking explicitly into account the modifications to the
Poisson equation.

Our formalism is also capable of describing the cosmological expansion
histories that are generated by the phenomenological XCDM
model~\cite{1997PhRvD..56.4439T}. This is demonstrated in
Fig.~\ref{fig:w0}, for a flat universe, where the values of the
equivalent equation-of-state parameter $w$ are shown as a function of
the perturbative parameter $\mu^4$, the matter density, and the value
of the cosmological constant.  In a similar manner, our framework can
also describe the cosmological expansion generated by the model
suggested by the Dark Energy Task Force~\cite{Albrecht:2006p1174}, in
which the equation-of-state parameter is allowed to change in time as
$w_0+(1-a)w_a$, since the values of the equivalent parameter $w$
calculated here depend indeed on cosmic time (cf.\ Eq.~\ref{wnumbs}). In
contrast to the phenomenological model, however, our framework offers
additional insight as to how the resulting gravity modifications agree
with structure formation observations, or with ``fifth-force''
constraints from other arenas.


We will study the behavior of the cosmological expansion predicted by
gravity with perturbative constraints for a Universe with a finite
spatial curvature as well as compare these predictions directly to
observations in forthcoming papers.

\acknowledgements

A.\,C.\ thanks the members of the Kavli Institute for Cosmological
Physics for their hospitality. S.\,D. thanks the Perimeter Institute (Canada)
for hospitality, and Wayne Hu, Justin Khoury and members of the Perimeter Institute
cosmology group for helpful discussions. We thank \'Eanna Flanagan for helpful comments. We also thank Feryal \"Ozel for carefully
reading the manuscript. This work was supported in part by an NSF
CAREER award at the University of Arizona.

\section{Appendix: Perturbative solution to 4th order for $\Lambda \ne 0$}

As discussed in the main text, perturbations to $\Lambda$CDM at order $\mu^4$ has closed-form solution. In particular: 
\begin{eqnarray}
a^{(4)}(t)&=&\left(\frac{\rho_{m,0}}{\Lambda}+4 \right)^2
\nonumber\\
&&
\left\{\frac{2\tan^{-1}\left[ \sqrt{3}\tanh \left( \frac{\sqrt{3\Lambda}}{2}t\right)\right]-9\sqrt{3}t}{96\sqrt{3}\tanh \left( \frac{\sqrt{3\Lambda}}{2}t\right)}\right.\nonumber \\
&&+\left.\frac{17-23\cosh(\sqrt{3\Lambda}t)+8\cosh(2\sqrt{3\Lambda}t)}{48\left[1-2\cosh(\sqrt{3\Lambda}t)\right]^2}\right\}\; \nonumber.
\end{eqnarray}

\end{document}